# Gödel and Physics

John D. Barrow
DAMTP
Centre for Mathematical Sciences
Cambridge University
Wilberforce Rd.,
Cambridge CB3 0WA
UK

Summary

We introduce some early considerations of physical and mathematical impossibility as preludes to Gödel's incompleteness theorems. We consider some informal aspects of these theorems and their underlying assumptions and discuss some the responses to these theorems by those seeking to draw conclusions from them about the completability of theories of physics. We argue that there is no reason to expect Gödel incompleteness to handicap the search for a description of the laws of Nature, but we do expect it to limit what we can predict about the outcomes of those laws, and examples are given. We discuss the 'Gödel universe', a solution of Einstein's equations describing a rotating universe where time travel is possible, which was discovered by Gödel in 1949, and the role it played in exposing the full spectrum of possibilities that a global understanding of space-time would reveal. Finally, we show how recent studies of supertasks have shown how global space-time structure determines the ultimate capability of computational devices within them.

## 1: Some Historical Background

### Physical Impossibilities

There is a long history of scientific and philosophical consideration of physical impossibilities[1]. The Aristotelian worldview outlawed the possibility that physical infinities or local physical vacua could be created or observed[2]. During the Middle Ages, physicists devised ingenious thought experiments to try to imagine how Nature could be 'tricked' into allowing an instantaneous vacuum to form, and then arguing about how this possibility was stopped from occurring by natural processes or, if that failed, by the invocation of a Cosmic Censor, to prevent its appearance[3]. Chemistry had its own ongoing alchemical debate about the possibility or impossibility of making gold from baser metals, and engineering maintained an enduring attachment to the quest for a perpetual motion machines that only fully abated when the consequences of laws of thermodynamics were systematically understood during the nineteenth century. Subtle examples, like Maxwell's sorting demon, still remained until they were eventually fully exorcised by the application of the modern thermodynamic theory of computation in 1961[4].

### Mathematical Impossibilities

Mathematicians also occasionally considered the question of impossibility in the context of a several fundamental problems of arithmetic, geometry, and algebra. Supposedly, in about 550BC, the Pythagoreans first encountered the 'irrationality' of numbers like √2 which cannot be expressed as the ratio of two integers ('irrational' originally meaning simply 'not a ratio', rather than beyond reason, as might be suggested today).[5]. Legend has it that this discovery was such a scandal, that the discoverer, Hippasos, was drowned by the members of the Pythagorean brotherhood for his trouble. This gives us the first glimpse of operations and questions which have no answers given a particular set of rules. In the first quarter of the 19$^{th}$ century, the problem of finding an explicit form for the solution of a general quintic algebraic equation in terms of its coefficients was proved to have no solution involving ordinary arithmetic operations and radicals by the young Norwegian mathematician, Henrik Abel[6]. Unlike the case of quadratic, cubic, or quartic equations, the general quintic cannot be solved by any exact formula. Just a few years later, in 1837, rigorous proofs were given that an angle of 60 degrees could not be trisected just by use of a straight edge and pair of compasses. These examples revealed for the first time, to those who looked at them in the right way, some hints as to the limitations of particular axiomatic systems.

In the light of the ongoing impact of Gödel's work on speculations about the limitations of the human mind, it is interesting to reflect briefly on the sociological and psychological effects of some of these early results. The existence of irrational numbers was of the deepest concern to the Pythagoreans; however, as far as we can judge, there were no deep philosophical questions about the limitations of mathematical reasoning raised by the demonstration that the quintic could not be solved. Yet, there was a change. Previously, there were many things thought impossible that could not be so proven despite many efforts to do so. But now there were proofs that something could not be done.

**Axiomatics**

The development of understanding of what constructions and proofs could be carried out by limited means, such as ruler and compass construction, or using only arithmetic operations and radicals, showed that axioms mattered. The power and scope of a system of axioms determined what its allowed rules of reasoning could encompass.

Until the 19th century, the archetypal axiomatic system was that of Euclidean geometry. But it important to appreciate that this system was not then viewed, as it is today, as just one among many axiomatic possibilities. Euclidean geometry was how the world really was. It was part of the absolute truth about the Universe. This gave it a special status and its constructions and elucidation, largely unchanged for more than 2000 years, provided a style that was aped by many works of philosophy and theology. The widespread belief in its absolute truth provided an important cornerstone for the beliefs of theologians and philosophers that human reason could grasp something of the ultimate nature of things. If challenged that this was beyond the power of our minds to penetrate, they could always point to Euclidean geometry as a concrete example of how and where this type of insight into the ultimate nature of things had already been possible. As a result, the discoveries, by Bolyai, Lobachevskii, Gauss, and Riemann, that other geometries existed, but in which Euclid's parallel postulate was not included, had a major impact outside of mathematics[7]. The existence of other logically consistent geometries meant that Euclid's geometry was not *the* truth: it was simply a model for some parts of the truth. As a result, new forms of relativism sprang up, nourished by the demonstration that even Euclid's ancient foundational system was merely one of many possible geometries – and indeed one of these alternatives was a far more appropriate model for describing the geometry of the Earth's surface than Euclid's. Curious books appeared about non-Euclidean models of government and economics. 'Non-Euclidean' became a byword for new and relative truth, the very latest intellectual fashion[8]. Later, new logics would be created as well, by changing the axioms of the classical logical system that Aristotle had defined.

Out of these studies emerged a deeper appreciation of the need for axioms to be consistently defined and clearly stated. The traditional realist view of mathematics as a description of how the world 'was' had to be superseded by a more sophisticated view that recognised mathematics to be an unlimited system of patterns which arise from the infinite number of possible axiomatic systems that can be defined. Some of those patterns appear to be made use of in Nature, but most are not. Mathematical systems like Euclidean geometry had been assumed to be part of the absolute truth about the world and uniquely related to reality. But the development of non-Euclidean geometries and non-standard logics meant that mathematical existence now meant nothing more than logical self-consistency (ie it must not be possible to prove that 0 = 1). It no longer had any necessary requirement of physical existence.

**Hilbert's programme**

The careful study of axiomatic systems revealed that even Euclid's beautiful development of plane geometry made use of unstated axioms. In 1882, Moritz Pasch gave a very simple example of an intuitively 'obvious' property of points and lines that could not be proved from Euclid's classical axioms[9]. If the points A, B, C, and D lie on a straight line such that B lies between A and C and C lies between B and D then it is not possible to prove that B lies between A and D. The picture of the set up made it appear inevitable but that is not a substitute for a proof.

A B C D

Pasch wanted to distinguish between the logical consequences of the axioms of geometry and those properties that we just assumed were intuitively true. For him, mathematical argumentation should not depend on any physical interpretation or visualisation of the quantities involved. He was concerned that axiomatic systems should be complete and has been described as 'the father of rigor in geometry' by Freudenthal[10].

David Hilbert, the greatest mathematician of the day, felt the influence of Pasch's writings both directly and through their effects on Peano's work[11] from 1882 to 1899, and began a systematic programme in 1899 to place mathematics upon a formal axiomatic footing[12]. This was a new emphasis, conveyed by Hilbert's remark that in mathematics 'One must be able to say .. -- instead of points, straight lines and planes -- tables, chairs, and beer mugs'[13]. He believed that it would be possible to determine the axioms underlying each part of mathematics (and hence of the whole), demonstrate that these axioms are self-consistent, and then show that the resulting system of statements and deductions formed from these axioms is both complete and decidable. More precisely, a system is *consistent* if we cannot prove that a statement S and its negation, ~S, are both true theorems. It is *complete* if for every statement S we can form in its language, either S or its negation, ~S is a true theorem. It is *decidable* if for every statement S that can be formed in its language, we can prove whether S is true or false. Thus, if a system is decidable it must be complete.

Hilbert's formalistic vision of mathematics was of a tight web of deductions spreading out with impeccable logical connections from the defining axioms. Indeed, mathematics was *defined* to be the collection of all those deductions. Hilbert set out to complete this formalisation of mathematics with the help of others, and believed that it would ultimately be possible to extend its scope to include sciences like physics[14] which were built upon applied mathematics. He began with Euclidean geometry and succeeded in placing it on a rigorous axiomatic basis. His programme then imagined strengthening the system by adding additional axioms, showing at each step that consistency and decidability remained, until eventually the system had become large enough to encompass the whole of arithmetic.

Hilbert's programme began confidently and he believed that it would just be a matter of time before all of mathematics was corralled within its formalistic web. Alas, the world was soon turned upon its head by the young Kurt Gödel. Gödel had completed one of the early steps in Hilbert's programme as part of his doctoral thesis, by proving the consistency and completeness of 1$^{st}$-order logic (later Alonzo Church and Alan Turing would show that it was not decidable). But the next steps that he took have ensured his fame as the greatest logician of modern times. Far from extending Hilbert's programme to achieve its key objective – a proof of the completeness of arithmetic – Gödel proved that any system rich enough to contain arithmetic must be incomplete and undecidable.

This took almost everyone by surprise, including John von Neumann, who was present at the conference in Königsberg (Hilbert's home town) on 7th September 1930 when Gödel briefly communicated his results, and quickly appreciated them -- even extending them, only to find that Gödel had already made the extension in a separate paper -- and Paul Finsler (who tried unsuccessfully to convince Gödel that he had discovered these results before him), and effectively killed Hilbert's programme with one stroke.

| Theory | Is it Consistent? | Is it Complete? | Is it Decidable? |
|---|---|---|---|
| Propositional calculus | Yes | Yes | Yes |
| Euclidean geometry | Yes | Yes | Yes |
| 1st order logic | Yes | Yes | No |
| Arithmetic (+,-) only | Yes | Yes | Yes |
| Arithmetic In full (+,-,×,÷) | ?? | No | No |

Table 1: Summary of the results established about the consistency, completeness, and decidability of simple logical systems.

## 2: Some Mathematical Jujitsu

### The Optimists and the Pessimists

Gödel 's monumental demonstration, that systems of mathematics have limits, gradually infiltrated the way in which philosophers and scientists viewed the world and our quest to understand it. Some commentators claimed that it shows that all human investigations of the Universe must be limited. Science is based on mathematics; mathematics cannot discover all truths; therefore science cannot discover all truths. One of Gödel's contemporaries, Hermann Weyl, described Gödel's discovery as exercising[15] 'a constant drain on the enthusiasm and determination with which I pursued my research work'. He believed that this underlying pessimism, so different from the rallying cry which Hilbert had issued to mathematicians in 1900, was shared 'by other mathematicians who are not indifferent to what their scientific endeavours mean in the context of man's whole caring and knowing, suffering and creative existence in the world'. In more recent times, one writer on theology and science with a traditional Catholic stance, Stanley Jaki, believes that Gödel's theorem prevents us from gaining an understanding of the cosmos as a necessary truth,

> 'Clearly then no scientific cosmology, which of necessity must be highly mathematical, can have its proof of consistency within itself as far as mathematics goes. In the absence of such consistency, all mathematical models, all theories of elementary particles, including the theory of quarks and gluons...fall inherently short of being that theory which shows in virtue of its a priori

> truth that the world can only be what it is and nothing else. This is true even if the theory happened to account with perfect accuracy for all phenomena of the physical world known at a particular time.' [16]

It constitutes a fundamental barrier to understanding of the Universe, for:

> 'It seems that on the strength of Gödel's theorem that the ultimate foundations of the bold symbolic constructions of mathematical physics will remain embedded forever in that deeper level of thinking characterized both by the wisdom and by the haziness of analogies and intuitions. For the speculative physicist this implies that there are limits to the precision of certainty, that even in the pure thinking of theoretical physics there is a boundary...An integral part of this boundary is the scientist himself, as a thinker..' [17]

Intriguingly, and just to show the important role human psychology plays in assessing the significance of limits, other scientists, like Freeman Dyson, acknowledge that Gödel places limits on our ability to discover the truths of mathematics and science, but interpret this as ensuring that science will go on forever. Dyson sees the incompleteness theorem as an insurance policy against the scientific enterprise, which he admires so much, coming to a self-satisfied end; for[18]

> 'Gödel proved that the world of pure mathematics is inexhaustible; no finite set of axioms and rules of inference can ever encompass the whole of mathematics; given any set of axioms, we can find meaningful mathematical questions which the axioms leave unanswered. I hope that an analogous situation exists in the physical world. If my view of the future is correct, it means that the world of physics and astronomy is also inexhaustible; no matter how far we go into the future, there will always be new things happening, new information coming in, new worlds to explore, a constantly expanding domain of life, consciousness, and memory.'

Thus, we see epitomised the optimistic and the pessimistic responses to Gödel. The ‘optimists’, like Dyson, see his result as a guarantor of the never-ending character of human investigation. They see scientific research as part of an essential part of the human spirit which, if it were completed, would have a disastrous de-motivating effect upon us – just as it did upon Weyl. The ‘pessimists’, like Jaki, Lucas[19], and Penrose[20], by contrast, interpret Gödel as establishing that the human mind cannot know all (maybe not even most) of the secrets of Nature.

Gödel's own view was as unexpected as ever. He thought that intuition, by which we can 'see' truths of mathematics and science, was a tool that would one day be valued just as formally and reverently as logic itself,

> 'I don't see any reason why we should have less confidence in this kind of perception, i.e., in mathematical intuition, than in sense perception, which induces us to build up physical theories and to expect that future sense perceptions will agree with them and, moreover, to believe that a question not decidable now has meaning and may be decided in the future.'[21]

However, it is easy to use Gödel’s theorem in ways that play fast and loose with the underlying assumptions of his theorem. Many speculative applications can be found spanning the fields of philosophy, theology, and computing and they have been examined in a lucid critical fashion by the late Torkel Franzén[22].

Gödel was not minded to draw any strong conclusions for physics from his incompleteness theorems. He made no connections with the Uncertainty Principle of quantum mechanics, which was advertised as another great deduction which limited our ability to know, and which was discovered by Heisenberg just a few years before Gödel made his discovery. In fact, Gödel was rather hostile to any consideration of quantum mechanics at all. Those who worked at the same Institute for Advanced Study (no one really worked *with* Gödel) believed that this was a result of his frequent discussions with Einstein who, in the words of John Wheeler (who knew them both) 'brainwashed Gödel' into disbelieving quantum mechanics and the Uncertainty Principle. Greg Chaitin records[23] this account of Wheeler's attempt to draw Gödel out on the question of whether there is a connection between Gödel incompleteness and Heisenberg’s Uncertainty Principle,

> 'Well, one day I was at the Institute for Advanced Study, and I went to Gödel's office, and there was Gödel. It was winter and Gödel had an electric heater and had his legs wrapped in a blanket. I said 'Professor Gödel, what connection do you see between your incompleteness theorem and Heisenberg's uncertainty principle?' And Gödel got angry and threw me out of his office!' [24]

The claim that mathematics contains unprovable statements -- physics is based on mathematics -- therefore physics will not be able to discover everything that is true, has been around for a long time. More sophisticated versions of it have been constructed which exploit the possibility of uncomputable mathematical operations being required to make predictions about observable quantities. From this vantage point, Stephen Wolfram, has conjectured that[25]

> 'One may speculate that undecidability is common in all but the most trivial physical theories. Even simply formulated problems in theoretical physics may be found to be provably insoluble.'

Indeed, it is known that undecidability is the rule rather than the exception amongst the truths of arithmetic[26].

**Drawing the Line Between Completeness and Incompleteness**

With these worries in mind, let us look a little more closely at what Gödel's result might have to say about physics. The situation is not so clear-cut as some commentators would often have us believe. It is useful to layout the precise assumptions that underlie Gödel's deduction of incompleteness. Gödel's theorem says that if a formal system is

1. *finitely specified*
2. *large enough to include arithmetic*
3. *consistent*

then it is *incomplete*.

Condition 1 means that there are not an uncomputable infinity of axioms. We could not, for instance, choose our system to consist of all the true statements about arithmetic because this collection cannot be finitely listed in the required sense.

Condition 2 means that the formal system includes all the symbols and axioms used in arithmetic. The symbols are 0, 'zero', S, 'successor of', +, ×, and =. Hence, the number two is the successor of the successor of zero, written as the *term* SS0, and two and plus two equals four is expressed as SS0 + SS0 = SSSS0.

The structure of arithmetic plays a central role in the proof of Gödel's theorem. Special properties of numbers, like their factorisations and the fact that any number can be factored in only one way as the product of prime divisors (eg. 130 = 2×5×13), were used by Gödel to establish a crucial correspondence between statements of mathematics and statements about mathematics. Thereby, linguistic paradoxes like that of the 'liar' could be embedded, like Trojan horses, within the structure of mathematics itself. Only logical systems which are rich enough to include arithmetic allow this incestuous encoding of statements about themselves to be made within their own language.

Again, it is instructive to see how these requirements might fail to be met. If we picked a theory that consisted of references to (and relations between) only the first ten numbers (0,1,2,3,4,5,6,7,8,9) with arithmetic modulo 10, then Condition 2 fails and such a mini-arithmetic is complete. Arithmetic makes statements about individual numbers, or terms (like SS0, above). If a system does not have individual terms like this but, like Euclidean geometry, only makes statements about a continuum of points, circles, and lines, in general, then it cannot satisfy Condition 2. Accordingly, as Alfred Tarski first showed, Euclidean geometry is complete. There is nothing magical about the flat, Euclidean nature of the geometry either: the non-Euclidean geometries on curved surfaces are also complete. Completeness can be long-winded though. A statement of geometry involving n symbols can take up to exp[exp[n]] computational steps to have its truth or falsity determined[27]. For just n = 10, this number amounts to a staggering $9.44 \times 10^{9565}$; for comparison, there have only been about $10^{27}$ nanoseconds since the apparent beginning of the Universe’s expansion history.

Similarly, if we had a logical theory dealing with numbers that only used the concept of 'greater than', without referring to any specific numbers, then it would be complete: we can determine the truth or falsity of any statement about real numbers involving just the 'greater than' relationship.

Another example of a system that is smaller than arithmetic is arithmetic without the multiplication, ×, operation. This is called Presburger[28] arithmetic (the full arithmetic is called Peano arithmetic after the mathematician who first expressed it axiomatically, in 1889). At first this sounds strange, in our everyday encounters with multiplication it is nothing more than a shorthand way of doing addition (eg 2+2+2+2+2+2 = 2×6), but in the full logical system of arithmetic, in the presence of logical quantifiers like 'there exists' or 'for any', multiplication permits constructions which are not merely equivalent to a succession of additions.

Presburger arithmetic is complete: all statements about the addition of natural numbers can be proved or disproved; all truths can be reached from the axioms[29]. Similarly, if we create another truncated version of arithmetic, which does not have addition, but retains multiplication, this is also complete. It is only when addition and multiplication are simultaneously present that incompleteness emerges. Extending the system further by adding extra operations like exponentiation to the repertoire of basic operations makes no difference. Incompleteness remains but no intrinsically new form of it is found. Arithmetic is the watershed in complexity.

The use of Gödel to place limits on what a mathematical theory of physics (or anything else) can ultimately tell us seems a fairly straightforward consequence. But as one looks more carefully into the question, things are not quite so simple. Suppose, for the moment, that all the conditions required for Gödel's theorem to hold are in place. What would incompleteness look like in practice? We are familiar with the situation of having a physical theory which makes accurate predictions about a wide range of observed phenomena: we might call it 'the standard model'. One day, we may be surprised by an observation about which it has nothing to say. It cannot be accommodated within its framework. Examples are provided by some so called ‘grand unified theories’ in particle physics. Some early editions of these theories had the property that all neutrinos must have zero mass. Now if a neutrino is observed to have a non-zero mass (as experiments have now confirmed) then we know that the new situation cannot be accommodated within our original theory. What do we do? We have encountered a certain sort of incompleteness, but we respond to it by extending or modifying the theory to include the new possibilities. Thus, in practice, incompleteness looks very much like inadequacy in a theory. It would become more like Gödel incompleteness if we could find no extension of the theory that could predict the new observed fact.

An interesting example of an analogous dilemma is provided by the history of mathematics. During the sixteenth century, mathematicians started to explore what happened when they added together infinite lists of numbers. If the quantities in the list get larger then the sum will 'diverge', that is, as the number of terms approaches infinity so does the sum. An example is the sum

$$1 + 2 + 3 + 4 + 5 + \ldots\ldots = \text{infinity}.$$

However, if the individual terms get smaller and smaller sufficiently rapidly[30] then the sum of an infinite number of terms can get closer and closer to a finite limiting value which we shall call the sum of the series; for example

$$1 + 1/9 + 1/25 + 1/36 + 1/49 + \ldots = \pi^2/8 = 1.2337005..$$

This left mathematicians to worry about a most peculiar type of unending sum,

$$1 - 1 + 1 - 1 + 1 - 1 + 1 - \ldots = ?????$$

If you divide up the series into pairs of terms it looks like (1-1) + (1-1) +.... and so on. This is just 0 + 0 + 0 +... = 0 and the sum is zero. But think of the series as 1-{1-1 + 1 - 1 + 1 -...} and it looks like 1 - {0} = 1. We seem to have proved that 0 = 1.

Mathematicians had a variety of choices when faced with ambiguous sums like this. They could reject infinities in mathematics and deal only with finite sums of numbers, or, as Cauchy showed in the early nineteenth century, the sum of a series like the last one must be defined by specifying more closely what is meant by its sum. The limiting value of the sum must be specified together with the procedure used to calculate it. The contradiction 0 = 1 arises only when one omits to specify the procedure used to work out the sum. In both cases it is different, and so the two answers are not the same because they arise in different axiomatic systems. Thus, here we see a simple example of how a limit is side-stepped by enlarging the concept which seems to create limitations.

Divergent series can be dealt with consistently so long as the concept of a sum for a series is suitably extended[31].

Another possible way of evading Gödel's theorem is if the physical world only makes use of the decidable part of mathematics. We know that mathematics is an infinite sea of possible structures. Only some of those structures and patterns appear to find existence and application in the physical world. It may be that they are all from the subset of decidable truths. Things may be even better protected than that: perhaps only computable patterns are instantiated in physical reality?

It is also possible that the conditions required to prove Gödel incompleteness do not apply to physical theories. Condition 1 requires the axioms of the theory to be listable. It might be that the laws of physics are not listable in this predictable sense. This would be a radical departure from the situation that we think exists, where the number of fundamental laws is believed to be not just listable, but finite (and very small). But it is always possible that we are just scratching the surface of a bottomless tower of laws, only the top of which has significant effects upon our experience. However, if there were an unlistable infinity of physical laws then we would face a more formidable problem than that of incompleteness.

An equally interesting issue is that of finiteness. It may be that the universe of physical possibilities is finite, although astronomically large[32]. However, no matter how large the number of primitive quantities to which the laws refer, so long as they are finite, the resulting system of inter-relationships will be complete. We should stress that although we habitually assume that there is a continuum of points of space and time this is just an assumption that is very convenient for the use of simple mathematics. There is no deep reason to believe that space and time are continuous, rather than discrete, at their most fundamental microscopic level; in fact, there are some theories of quantum gravity that assume that they are not. Quantum theory has introduced discreteness and finiteness in a number of places where once we believed in a continuum of possibilities. Curiously, if we give up this continuity, so that there is not necessarily another point in between any two sufficiently close points you care to choose. Space-time structure becomes infinitely more complicated because continuous functions can be defined by their values on the rationals. Many more things can happen. This question of finiteness might also be bound up with the question of whether the universe is finite in volume and whether the number of elementary particles (or whatever the most elementary entities might be) of Nature are finite or infinite in number. Thus, there might only exist a finite number of terms to which the ultimate logical theory of the physical world applies. Hence, it would be complete.

A further possibility with regard to the application of Gödel to the laws of physics is that Condition 2 of the incompleteness theorem might not be met. How could this be? Although we seem to make wide use of arithmetic, and much larger mathematical structures, when we carry out scientific investigations of the laws of Nature, this does not mean that the inner logic of the physical Universe needs to employ such a large structure. It is undoubtedly convenient for us to use large mathematical structures together with concepts like infinity but this may be an anthropomorphism. The deep structure of the Universe may be rooted in a much simpler logic than that of full arithmetic, and hence be complete. All this would require would be for the underlying structure to contain either addition or multiplication but not both. Recall that all the sums that you have ever done have used multiplication simply as a shorthand for addition. They would be possible in Presburger arithmetic as well. Alternatively, a basic structure of reality that made use of simple relationships of a geometrical variety, or which derived from 'greater than' or 'less than' relationships, or subtle combinations of them all could also remain complete[33]. The fact that Einstein's theory of general relativity replaces many physical notions like force and weight by *geometrical* distortions in the fabric of space-time may well hold some clue about what is possible here.

There is a surprisingly rich range of possibilities for a basic representation of mathematical physics in terms of systems which might be decidable or undecidable. Tarski showed that, unlike Peano’s arithmetic of addition and multiplication of natural numbers, the first-order theory of real numbers under addition and multiplication is decidable. This is rather surprising and may give some hope that theories of physics based on the reals or complex numbers will evade undecidability in general. Tarski also went on to show that many mathematical systems used in physics, like lattice theory, projective geometry, and Abelian group theory are decidable, while others, notably non-Abelian group theory are not[34]. Little consideration seems to be have given to the consequences of these results to the development of ultimate theories of physics.

There is another important aspect of the situation to be keep in view. Even if a logical system is complete, it always contains unprovable 'truths'. These are the axioms which are chosen to define the system. And after they are chosen, all the logical system can do is deduce conclusions from them. In simple logical systems, like Peano arithmetic, the axioms seem reasonably obvious because we are thinking backwards -- formalising something that we have been doing intuitively for thousands of years. When we look at a subject like physics, there are parallels and differences. The axioms, or laws, of physics are the prime target of physics research. They are by no means intuitively obvious, because they govern regimes that can lie far outside of our experience. The outcomes of those laws are unpredictable in certain circumstances because they involve symmetry breakings. Trying to deduce the laws from the outcomes is not something that we can ever do uniquely and completely by means of a computer programme.

Thus, we detect a completely different emphasis in the study of formal systems and in physical science. In mathematics and logic, we start by defining a system of axioms and laws of deduction. Then, we might try to show that the system is complete or incomplete, and deduce as many theorems as we can from the axioms. In science, we are not at liberty to pick any logical system of laws that we choose. We are trying to find the system of laws and axioms (assuming there is one -- or more than one perhaps) that will give rise to the outcomes that we see. As we stressed earlier, it is always possible to find a system of laws which will give rise to any set of observed outcomes. But it is the very set of unprovable statements that the logicians and the mathematicians ignore -- the axioms and laws of deduction -- that the scientist is most interested in discovering rather than simply assuming. The only hope of proceeding as the logicians do, would be if for some reason there is only one possible set of axioms or laws of physics. So far, this does not seem likely[35]; even if it were we would not be able to prove.

## 3: Laws versus Outcomes

### Symmetry Breaking

The structure of modern physics presents with an important dichotomy. It is important to appreciate this division in order to understand the significance of Gödel incompleteness for physics. The fundamental laws of Nature governing the weak, strong, electromagnetic, and gravitational forces, are all local gauge theories derived from the maintenance of particular mathematical symmetries. As these forces become unified, the number of symmetries involved will be reduced until ultimately (perhaps) there is only one over-arching symmetry dictating the form the laws of Nature – a so-called 'Theory of Everything', of which M theory is the current candidate. Thus the laws of Nature are in a real sense 'simple' and highly symmetrical. The ultimate symmetry which unites them must possess a number of properties in order to accommodate all the low-energy manifestations of the separate forces, the states that look like elementary 'particles' with all their properties, and it must be big enough for them all to fit in.

There is no reason why Gödel incompleteness should hamper the search for this all-encompassing symmetry governing the *laws* of Nature. This search is, at root, a search for a pattern, perhaps a group symmetry or some other mathematical prescription. It need not be complicated and it probably has a particular mathematical property that makes it specially (or even uniquely) fitted for this purpose.

In reality, we never 'see' laws of Nature: they inhabit a Platonic realm. Rather, we witness their *outcomes*. This is an important distinction, because the outcomes are quite different to the laws that govern them. They are asymmetrical and complicated and need possess none of the symmetries displayed by the laws. This is fortunate because, were it not so, we could not exist. If the outcomes of the laws of Nature possessed all the symmetries of the laws then nothing could happen which did not respect them. There could be no structures located at particular times and places, no directional asymmetries, and nothing happening at any one moment. All would be unchanging and empty.

This dichotomy between laws and outcomes is what I would call 'the secret of the universe'. It is what enables a Universe to be governed by a very small number (perhaps just one) of simple and symmetrical laws, yet give rise to an unlimited number of highly complex, asymmetrical states – of which we are one variety[36].

Thus, whereas there is no reason to worry about Gödel incompleteness frustrating the search for the mathematical descriptions of the laws of Nature, we might well expect Gödel incompleteness to arise in our attempts to describe some of the complicated sequences of events that arise as outcomes to the laws of Nature.

### Undecidable Outcomes

Specific examples have been given of physical problems in which the outcomes of their underlying laws are undecidable. As one might expect from what has just been said, they do not involve an inability to determine something fundamental about the nature of the laws of physics, or even the most elementary particles of matter. Rather, they involve an inability to perform some specific mathematical calculation, which inhibits our ability to determine the course of events in a well-defined physical problem. However, although the problem may be mathematically well defined, this does not mean that it is possible to create the precise conditions required for the undecidability to exist.

An interesting series of examples of this sort have been created by the Brazilian mathematicians, Francisco Doria and N. da Costa[37]. Responding to a challenge problem posed by the Russian mathematician Vladimir Arnold, they investigated whether it was possible to have a general mathematical criterion which would decide whether or not any equilibrium was stable. A stable equilibrium is a situation like a ball sitting in the bottom of a basin – displace it slightly and it returns to the bottom; an unstable equilibrium is like a needle balanced vertically – displace it slightly and it moves away from the vertical[38]. When the equilibrium is of a simple nature this problem is very elementary; first-year science students learn about it. But, when the equilibrium exists in the face of more complicated couplings between the different competing influences, the problem soon becomes more complicated than the situation studied by science students[39]. So long as there are only a few competing influences the stability of the equilibrium can still be decided by inspecting the equations that govern the situation. Arnold's challenge was to discover an algorithm which tells us if this can always be done, no matter how many competing influences there are, and no matter how complex their inter-relationships. By 'discover' he meant find a formula into which you can feed the equations which govern the equilibrium along with your definition of stability, and out of which will pop the answer 'stable' or unstable'.

Da Costa and Doria discovered that there can exist no such algorithm. There exist equilibria characterised by special solutions of mathematical equations whose stability is undecidable. In order for this undecidability to have an impact on problems of real interest in mathematical physics the equilibria have to involve the interplay of very large numbers of different forces. While such equilibria cannot be ruled out, they have not arisen yet in real physical problems. Da Costa and Doria went on to identify similar problems where the answer to a simple question, like 'will the orbit of a particle become chaotic', is Gödel undecidable. They can be viewed as physically grounded examples of the theorems of Rice[40] and Richardson[41] which show, in a precise sense, that only trivial properties of computer programs are algorithmically decidable. Others have also tried to identify formally undecidable problems. Geroch and Hartle have discussed problems in quantum gravity that predict the values of potentially observable quantities as a sum of terms whose listing is known to be a Turing uncomputable operation[42]. Pour-El and Richards[43] showed that very simple differential equations, which are widely used in physics, like the wave equation, can have uncomputable outcomes when the initial data is not very smooth. This lack of smoothness gives rise to what mathematicians call an 'ill-posed' problem. It is this feature that gives rise to the uncomputability. However, Traub and Wozniakowski[44] have shown that every ill-posed problem is well-posed on the average under rather general conditions. Wolfram[45] gives examples of intractability an undecidability arising in condensed matter physics and even believes that undecidability is typical in physical theories.

The study of Einstein's' general theory of relativity also produces an undecidable problem if the mathematical quantities involved are unrestricted[46]. When one finds an exact solution of Einstein's equations it is always necessary to discover whether it is just another, known solution that is written in a different form. Usually, one can investigate this by hand, but for complicated solutions computers can help. For this purpose we require computers programmed to carry out algebraic manipulations. They can check various quantities to discover if a given solution is equivalent to one already sitting in its memory bank of known solutions. In practical cases encountered so far, this checking procedure comes up with a definite result after a small number of steps. But, in general, the comparison is an undecidable process equivalent to another famous undecidable problem of pure mathematics, 'the word problem' of group theory, first posed by Max Dehn[47] in 1911 and shown to be undecidable in 1955[48].

The tentative conclusions we should draw from this discussion is that, just because physics makes use of mathematics, it is by no means required that Gödel places any limit upon the overall scope of physics to understand the laws of Nature[49]. The mathematics that Nature makes use of may be smaller, and simpler than is needed for undecidability to rear its head.

## 4: Gödel and Space-time Structure

### Space-time in a Spin

Although Kurt Gödel is famous amongst logicians for his incompleteness theorems, he is also famous amongst cosmologists, but for a quite different reason. In 1949, inspired by his many conversations with Einstein about the nature of time and Mach's principle, Gödel found a new and completely unsuspected type of solution to Einstein's equations of general relativity[50]. Gödel's solution was a universe that rotates and permits time travel to occur into the past.

This was the first time that the possibility of time travel (into the past) had emerged in the context of a theory of physics. The idea of time travel first appeared in H.G. Wells' famous 1895 story, *The Time Machine*, but it was widely suspected that backwards[51] time travel into the past would in some way be in conflict with the laws of Nature. Gödel's universe showed that was not necessarily so: it could arise as a consequence of a theory obeying all the conservation laws of physics. Time travel into the future is a relatively uncontroversial matter and is just another way of describing the observed effects of time dilation in special relativity.

Gödel's universe is not the one that we live in. For one thing, Gödel's universe is not expanding; for another, there is no evidence that our Universe is rotating – and if it is, then its rate of spin must be at least $10^5$ times slower than its expansion rate because of the isotropy of the microwave background radiation[52]. Nonetheless, Gödel's universe was a key discovery in the study of space-time and gravitation. If time travel was possible, perhaps it could arise in other universes which are viable descriptions of our own?

But the influence of Gödel's solution on the development of the subject was more indirect. It revealed for the first time the subtlety of the *global* structure of space-time, particularly when rotation is present. Previously, the cosmological models that were studied tended to be spatially homogeneous with simple topologies and high degrees of global symmetry that ruled out or disguised global structure. Later, in 1965, Roger Penrose would apply powerful new methods of differential topology to this problem and prove the first singularity theorems in cosmology. The possibility of closed time-like curves in space-time that Gödel had revealed meant that specific vetos had to be included in some of these theorems to exclude their presence, otherwise past incompleteness of geodesics could be avoided by periodically reappearing in the future. It was Gödel's universe that first showed how unusual space-times could be whilst still remaining physically and factually consistent. Prior to its discovery, physicists and philosophers of science regarded time travel to the past as the necessary harbinger of factual contradictions. But Gödel's solution shows that there exist self-consistent histories which are periodic in space and time[53]. It continues to be studied as a key example of an intrinsically general-relativistic effect and its full stability properties have been elucidated only recently[54]. Some of its unusual properties are explained in the accompanying article by Wolfgang Rindler.

In recent years, Gödel’s study of space-time structure and his work on the incompleteness of logical systems have been pulled together in a fascinating way. It has been shown that there is a link between the global structure of a space-time and the sorts of computations that can be completed within them. This unexpected link arises from a strange old problem with a new name: is it possible to do an infinite number of things in a finite amount of time? And the new name for such a remarkable old activity is a ‘supertask’.

**Supertasks**

The ancients, beginning with Zeno, were challenged by the paradoxes of infinities on many fronts[55]. But what about philosophers today? What sort of problems do they worry about? There are live issues on the interface between science and philosophy that are concerned with whether it is possible to build an "infinity machine" that can perform an infinite number of tasks in a finite time. Of course, this simple question needs some clarification: what exactly is meant by "possible", "tasks", "number", "infinite", "finite" and, by no means least, by "time". Classical physics appears to impose few physical limits on the functioning of infinity machines because there is no limit to the speed at which signals can travel or switches can move. Newton's laws allow an infinity machine. This can be seen by exploiting a discovery about Newtonian dynamics made in 1971 by the US mathematician Jeff Xia[56]. First take four particles of equal mass and arrange them in two binary pairs orbiting with equal but oppositely-directed spins in two separate parallel planes, so the overall angular momentum is zero. Now, introduce a fifth much lighter particle that oscillates back and forth along a perpendicular line joining the mass centres of the two binary pairs. Xia showed that such a system of five particles will expand to infinite size in a *finite* time!

How does this happen? The little oscillating particle runs back and forth between the binary pairs, each time creating an unstable meeting of three bodies. The lighter particle then gets kicked back and the binary pair recoils outwards to conserve momentum. The lighter particle then travels across to the other binary and the same *ménage à trois* is repeated there. This continues without end, accelerating the binary pairs apart so strongly that they become infinitely separated while the lighter particle undergoes an infinite number of oscillations in the finite time before the system achieves infinite size.

Unfortunately (or perhaps fortunately), this behaviour is not possible when relativity is taken into account. No information can be transmitted faster than the speed of light and gravitational forces cannot become arbitrarily strong in Einstein's theory of motion and gravitation; nor can masses get arbitrarily close to each other and recoil - there is a limit to how close separation can get, after which an "event horizon" surface encloses the particles to form a black hole[57]. Their fate is then sealed – no such infinity machine could send information to the outside world.

But this does not mean that all relativistic infinity machines are forbidden[58]. Indeed, the Einsteinian relativity of time that is a requirement of all observers, no matter what their motion, opens up some interesting new possibilities for completing infinite tasks in finite time. Could it be that one moving observer could see an infinite number of computations occurring even though only a finite number had occurred according to someone else? Misner[59] and Barrow and Tipler[60] have shown that there are examples of entire universes in which an infinite number of oscillations occur on approach to singularities in space-time but it is necessary for the entire universe to hit the singularity; in effect, the whole universe is the infinity machine. It still remains to ask whether a local infinity machine could exist and send us signals as a result of completing an infinite number of operations in a finite amount of our time.

The famous motivating example of this sort of temporal relativity is the so-called 'twin paradox'. Two identical twins are given different future careers. Tweedlehome stays at home while Tweedleaway goes away on a space flight at a speed approaching that of light. When they are eventually reunited, relativity predicts that Tweedleaway will find Tweedlehome to be much older. The twins have experienced different careers in space and time because of the acceleration and deceleration that Tweedleaway underwent on his round trip.

So can we ever send a computer on a journey so extreme that it could accomplish an infinite number of operations by the time it returns to its stay-at-home owner? Itamar Pitowsky first argued[61] that if Tweedleaway could accelerate his spaceship sufficiently strongly, then he could record a finite amount of the universe's history on his own clock while his twin records an infinite amount of time on his clock. Does this, he wondered, permit the existence of a "Platonist computer" - one that could carry out an infinite number of operations along some trajectory through space and time and print out answers that we could see back home. Alas, there is a problem - for the receiver to stay in contact with the computer, he also has to accelerate dramatically to maintain the flow of information. Eventually the gravitational forces become stupendous and he is torn apart.

Not withstanding these problems a check-list of properties has been compiled for universes that can allow an infinite number of tasks to be completed in finite time, or "supertasks" as they have become known. These are called Malament-Hogarth (MH) universes after David Malament, a University of Chicago philosopher, and Mark Hogarth[62], a former Cambridge University research student, who, in 1992, investigated the conditions under which they were theoretically possible. Supertasks[63] open the fascinating prospect of finding or creating conditions under which an infinite number of things can be seen to be accomplished in a finite time. This has all sorts of consequences for computer science and mathematics because it would remove the distinction between computable and uncomputable operations. It is something of a surprise that MH universes (see Figure 1) are self-consistent mathematical possibilities but, unfortunately, have properties that suggest they are not realistic physical possibilities unless we embrace some disturbing notions, such as the prospect of things happening without causes, and travel backwards through time.

The most serious by-product of being allowed to build an infinity machine is rather more alarming though. Observers who stray into bad parts of these universes will find that being able to perform an infinite number of computations in a finite time also means that any amount of radiation, no matter how small, gets compressed to zero wavelength and amplified to infinite frequency and energy along the infinite computational trail. Thus any attempt to transmit the output from an infinite number of computations will zap the receiver and destroy him. So far, these dire problems seem to rule out the practicality of engineering a relativistic infinity machine in such a way that we could safely receive and store the information. But the universes in which infinite tasks are possible in finite time includes a type of space that plays a key role in the structure of the very superstring theories that looked so appealingly finite.

If you could see the output from an infinity machine that completes supertasks then you have the possibility of deciding undecidable problems by direct search through the infinite catalogue of possibilities: Turing's uncomputable operations seem to become completeable in a finite amount of our wristwatch time. Is this really possible? Remarkably, Hogarth showed[64] in some space-times it was possible to decide Gödel undecidable questions by direct search by sending a computer along a certain space-time path, $\gamma$, so that it could print out and send you the answer to the question. Now, create a hierarchy of n space-time structures of ascending complexities such that the $n^{th}$ in the sequence allows a supertask to be completed which can check the truth of any arithmetical assertion made in the $n^{th}$ but not the $(n+1)^{st}$ quantifier arithmetic in Kleene's logical hierarchy, by which logicians calibrate the complexity of possible logical expressions. There is a neat one-to-one correspondence between the list of space-times and the complexity of the logical statements that they can decide. Subsequently, Etesi and Nemeti[65] showed that some relations on natural numbers which are neither universal nor co-universal, can be decided in Kerr space-times.Welch[66] recently generalised these results to show that the computational capability of space-times could be raised beyond that of arithmetic to hyper-arithmetics, and showed that there is upper bound on the computational ability in any space-time which is a universal constant defined by the space-time.

Thus, in conclusion, we find that Gödel's ideas are still provoking new research programmes and unsuspected properties of the worlds of logical and physical reality. His incompleteness theorems should not be a drain on our enthusiasm to seek out and codify the laws of Nature: there is no reason for them to limit that search for the fundamental symmetries of Nature in any significant way. But, by contrast, in situations of sufficient complexity, we do expect to find that Gödel incompleteness places limits on our ability to use those laws to predict the future, carry out specific computations, or build algorithms: incompleteness besets the outcomes of very simple laws of Nature. Finally, if we study universes, then Gödel's impact will always be felt as we try to reconcile the simple local geometry of space and time with the extraordinary possibilities that its exotic global structure allows. Space-time structure defines what can be proved in a universe.

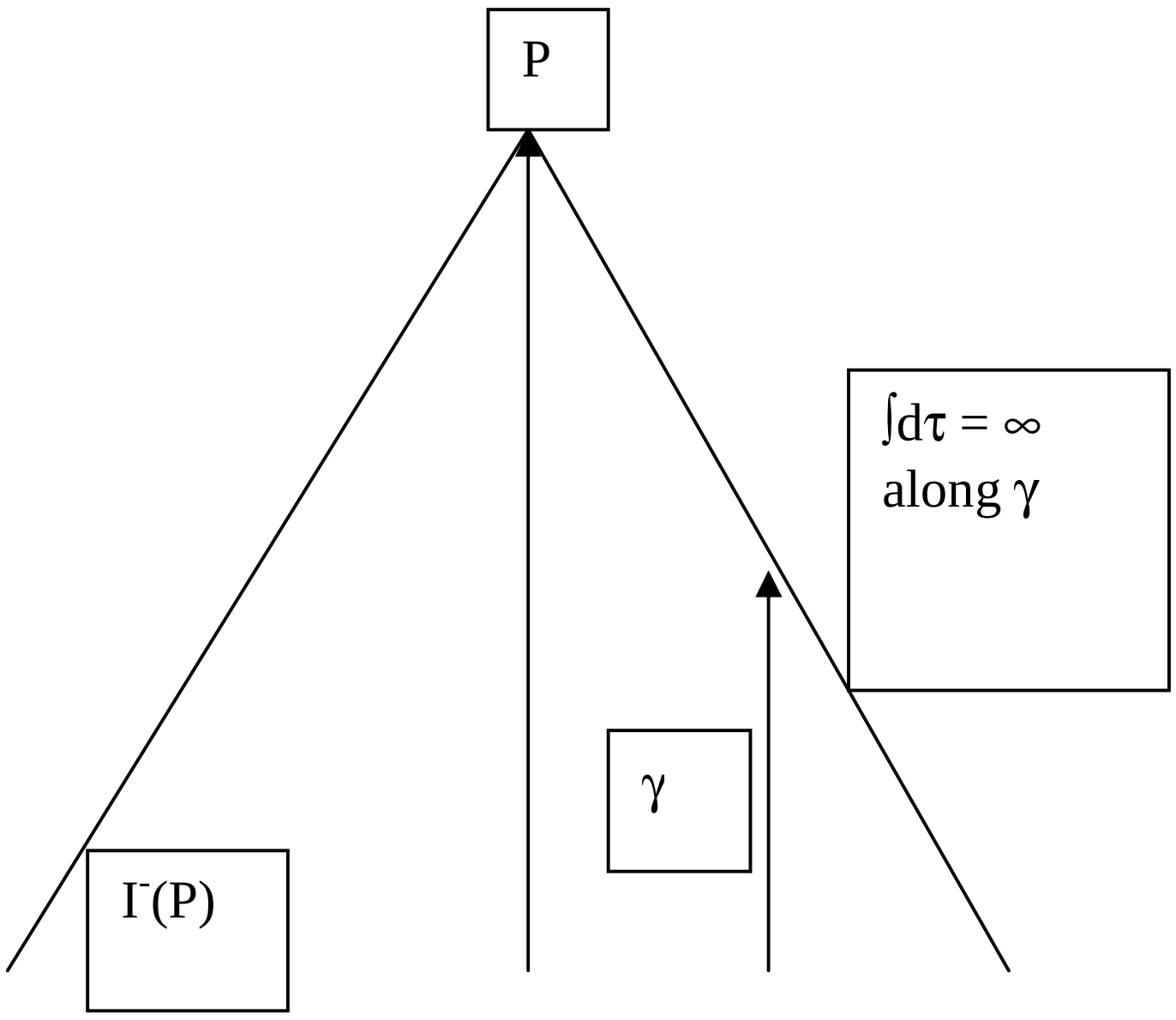

**Figure 1**: The space-time of an MH space-time with time mapped vertically and space (compressed to one dimension) horizontally. We are located at P and our causal past, $I^-(P)$ consists of all the events that can influence us. If there is a path in our past, $\gamma$, such that there is an infinite amount of its own time passing on approach to the space-time point where it intersects our past line cone, the edge of $I^-(P)$.

**Acknowledgements** I would like to thank Philip Welch for discussions and Christos Papadimitriou for many helpful suggestions.

---

[1] J.D. Barrow, *Impossibility*, Oxford UP, Oxford, (1998).
[2] J.D. Barrow, *The Book of Nothing*, Cape, London, (2000).
[3] E. Grant, *Much Ado About Nothing: Theories of Space and Vacuum from the Middle Ages to the Scientific Revolution*, Cambridge UP, Cambridge, (1981). Earlier examples of some of these challenging thought experiments can be found in Lucretius, *De Rerum Natura* Book 1.
[4] See H. Leff and A. Rex (eds.), *Maxwell's* Demon, Princeton UP, Princeton, (1990) for an overview with reprints of all the crucial papers.
[5] K. Guthrie (ed.), *The Pythagorean Sourcebook and Library*, Phares Press, Grand Rapids, (1987).
[6] P. Pesic, *Abel's Proof: an essay on the sources and meaning of mathematical unsolvability*, MIT Press, Cambridge (2003). This book also contains a new translation of Abel's 1824 discovery paper.
[7] J. Richards, The Reception of a Mathematical Theory: Non-Euclidean geometry in England 1868-1883, in *Natural Order: Historical Studies of Scientific Culture*, eds. B. Barnes and S. Shapin, Sage Publ., Beverly Hills, (1979).
[8] J.D. Barrow, *Pi in the Sky*, Oxford UP, Oxford, (1992), pp.8-15.
[9] M. Pasch, *Vorlesungen über neuere Geometrie*, Leipzig, (1882)

[10] H. Freudenthal, *The main trends in the foundation of geometry in the 19th century*. Stanford UP, Stanford, (1962), p.619, quoted in Kennedy op. cit. p.133.
[11] H.C. Kennedy, Amer. Math. Monthly 79, 133-6 (1972).
[12] See J. Gray, *The Hilbert Challenge*, Oxford UP, Oxford, (2000) which contains the text of Hilbert's Address to the International Congress of Mathematicians in 1900 on pp. 240-282 and M. Toepell, Archive for History of Exact Sciences 35, 329 (1986)
[13] D. Hilbert, quoted in C. Reid, *Hilbert*, Springer-Verlag, NY, (1970), p. 57.
[14] L. Corry, Archive for History of Exact Sciences 51, 83 (1997)
[15] H. Weyl, Amer. Math. Monthly 53, 13 (1946).
[16]S. Jaki, *Cosmos and Creator*, Scottish Academic Press, Edinburgh, (1980), p49.
[17]S. Jaki, *The Relevance of Physics,* Univ. Chicago P, Chicago, (1966), p129.
[18] F. Dyson, The World on a String, New York Review of Books, May 13th, (2004)
[19] J.R. Lucas, Philosophy 36, 112 (1961) and *Freedom of the Will* (1970).
[20] R. Penrose, *The Emperor's New Mind,* Oxford UP., Oxford, (19XX).
[21] K. Gödel, What is Cantor's Continuum Problem?, Philosophy of Mathematics ed. P. Benacerraf & H. Putnam p 483
[22] T. Frenzén, *Gödel's Theorem: An Incomplete Guide to its Use and Abuse,* A.K. Peters, Wellesley, (2005).
[23] Quoted in J. Bernstein, *Quantum Profiles*, Basic Bks., NY, (1991). Pp. 140-1; see also J.D. Barrow, *Impossibility,* Oxford UP, Oxford, (1998), chap. 8.
[24]Recorded by Chaitin see J. Bernstein, *Quantum Profiles*, pp 140-1 and K. Svozil, *Randomness and Undecidability in Physics*, World Scientific, Singapore, (1993), p112.
[25]S. Wolfram, *Cellular Automata and Complexity, Collected Papers,* Add. Wesley, Reading MA, (1994).
[26]C. Calude, *Information and randomness -- An Algorithmic Perspective,* Springer, Berlin, (1994); C. Calude, H. Jürgensen, and M. Zimand, Is Independence an Exception?, Appl. Math. Comput. 66, 63 (1994); K. Svozil, in *Boundaries and Barriers: on the limits of scientific knowledge,* eds. J. Casti and A. Karlqvist, Add. Wesley, NY, (1996), p. 215.
[27] D. Harel, *Computers Ltd.,* Oxford UP, Oxford (2000).
[28] M. Presburger, Comptes Rendus du Congrès de Mathématiciens des Pays Slaves, Warsaw pp. 92-101 (1929); D.C. Cooper, Theorem Proving in Arithmetic without Multiplication, in B. Meltzer and D. Michie (eds.), *Machine Intelligence,* Edinburgh UP, Edinburgh, (1972), pp. 91-100.
[29]The decision procedure is in general double-exponentially long, though. That is, the computational time required to carry out N operations grows as $(2^N)^N$. Presburger arithmetic allows us to talk about positive integers, and variables whose values are positive integers. If we enlarge it by permitting the concept of sets of integers to be used, then the situation becomes almost unimaginably intractable. It has been shown that this system does not admit even a K-fold exponential algorithm, for any finite K. The decision problem is said to be non-elementary in such situations. The intractability is unlimited; see M.J. Fischer and M.O. Rabin, Super-Exponential Complexity of Presburger Arithmetic, Proc. SIAM-AMS Symp. 7, 27-41 (1974).Appl. Maths, 7, 27-41 (1974).
[30]That the terms in the sum get progressively smaller is a necessary but not a sufficient condition for an infinite sum to be finite. For example, the sum 1+1/2+1/3+1/4+1/5+... is infinite.
[31]R. Rosen, On the Limitations of Scientific Knowledge, in *Boundaries and Barriers: on the limits of scientific knowledge,* eds. J.L. Casti and A. Karlqvist, Add. Wesley, NY, (1996), p. 199.
[32] This could be because it has positive spatial curvature or a compact spatial topology like that of the 3-torus even though it has non-positive curvature, see J.D. Barrow, *The Infinite Book*, Cape, London, (2005), chap. 7.
[33]John A. Wheeler has speculated about the ultimate structure of space-time being a form of 'pregeometry' obeying a calculus of propositions restricted by Gödel incompleteness. We are proposing that this pregeometry might be simple enough to be complete; see C. Misner, K. Thorne, and J.A. Wheeler, *Gravitation,* W.H. Freeman, San. Fran, (1973), pp. 1211-2.
[34] A. Tarski, A. Mostowski and R.M. Robinson, *Undecidable theories*, North Holland, Amsterdam, (1953).
[35]The situation in superstring theory is still very fluid. There appear to exist many different, logically self-consistent superstring theories, but there are strong indications that they may be different representations of a much smaller number (maybe even just one) theory.

[36] J.D. Barrow, *New Theories of Everything*, Oxford UP, Oxford, (2007).
[37]N.C. da Costa and F. Doria, Int. J. Theor. Phys. 30, 1041 (1991), Found. Phys. Letts., 4, 363 (1991)
[38]Actually, there are other more complicated possibilities clustered around the dividing line between these two simple possibilities and it is these that provide the indeterminacy of the problem in general.
[39] It is may not be determined at linear order if there are eigenvalues which are zero or are purely imaginary.
[40] H.G. Rice, Trans. Amer. Math. Soc. 74, 358 (1953).
[41] D. Richardson, Symbolic Logic 33, 514 (1968).
[42]R. Geroch and J. Hartle, Computability and Physical Theories, Foundations of Physics, 16, 533 (1986). The problem is that the calculation of a wave function for a cosmological quantity involves the sum of quantities evaluated on every 4-dimensional compact manifold in turn. The listing of this collections of manifolds is uncomputable.
[43]M.B. Pour-El and I. Richards, A computable ordinary differential equation which possesses no computable solution, Ann. Math. Logic, 17, 61 (1979), The wave equation with computable initial data such that its unique solution is not computable, Adv. Math. 39, 215 (1981), Non-computability in models of physical phenomena, Int. J. Theor. Phys. 21, 553 (1982).
[44]J.F. Traub and H. Wozniakowski, Information-based Complexity: New Questions for Mathematicians, The mathematical Intelligencer, 13, 34 (1991).
[45]S. Wolfram, Undecidability and Intractability in Theoretical Physics, Phys. Rev. Lett., 54, 735 (1985); Origins of Randomness in Physical Systems, Phys. Rev. Lett. 55, 449 (1985); Physics and Computation, Int. J. Theo. Phys. 21, 165 (1982).
[46]If the metric functions are polynomials then the problem is decidable, but is computationally double-exponential. If the metric functions are allowed to be sufficiently smooth then the problem becomes undecidable, see the article on Algebraic Simplification by Buchberger and Loos in Buchberger, Loos and Collins, *Computer Algebra: symbolic and algebraic computation'*, 2nd ed., Springer, Wien, (1983). I am grateful to Malcolm MacCallum for supplying these details.
[47] M. Dehn, Math. Ann. 71, 73 (1911).
[48] See W. Boone, Annals of Mathematics, 70, 207 (1959)
[49] For a fuller discussion of the whole range of limits that may exist on our ability to understand the universe see the author's discussion in J.D. Barrow, *Impossibility*, Oxford UP, (1998) on which the discussion here is based.
[50] K. Gödel, Rev. Mod. Phys. 21, 447 (1949). For details of the global structure of the Gödel universe see S.W. Hawking and G.F.R. Ellis, *The Large scale Structure of Space-time*, Cambridge UP, Cambridge (1973); for a study of its stability and a discussion of its Newtonian counterpart, see J.D. Barrow and C.G. Tsagas, Class. Quantum Gravity 21, 1773 (2004).
[51] Forward time travel into the future is routinely observed and is a simple consequence of special relativity. For example, in the case of the twin paradox, the twin who has undergone accelerations returns to find that he is younger than his twin: he has in effect time travelled into his twin's future.
[52] J.D. Barrow, R. Juszkiewicz and D.H. Sonoda, MNRAS 213, 917 (1985).
[53] J.D. Barrow, *The Infinite Book*, Jonathan Cape, London (2005).
[54] J.D. Barrow and C. Tsagas, Class. Quantum Gravity 21, 1773 (2004).
[55] A. Grünbaum, *Modern Science and Zeno's Paradoxes*, Allen and Unwin, London (1968), and *Philosophical Problems of Space and Time*, 2nd. Edn., Reidel, Dordrecht, (1973), chap. 18.
[56] Z. Xia, Ann. Math. 135,, 411 (1992) and D.G. Saari and Z. Xia, J. Diff. Eqns. 82, 342 (1989). The connection between Xia's remarkable construction is pointed out in and J.D. Barrow, *The Infinite Book*, Cape, London, (2005), chap.10. Note however, that the initial conditions needed for Xia's specific example are special (a Cantor set) and will not arise in practice.
[57] Once two masses, each of mass M, fall inside a radius $4GM/c^2$ an event horizon forms around them.
[58] We could be living inside a very large black hole along with the infinity machine.
[59] C.W. Misner, Phys. Rev. 186, 1328 (1969).
[60] J.D. Barrow and F.J. Tipler, *The Anthropic Cosmological Principle*, Oxford UP, Oxford (1986), chap. 10.
[61] I. Pitowsky, Iyyun 39, 81 (1990).
[62] M.L. Hogarth, Found. Phys. Lett. 5, 173 (1992).

[63] J. Earman and Norton, Phil. of Sci. 60, 22 (1993) and J.D. Barrow, *The Infinite Book*, Cape, London, (2005), chap.10.
[6464] M. Hogarth, Brit. Journal Phil. of Sci. 55, 681 (2004).
[65] G. Etesi and I. Németi, Int. J. Theor. Phys. 41, 341 (2002).
[66] P.D. Welch, gr-qc/0609035